\documentclass[12pt]{article}

\usepackage[utf8]{inputenc} 
\usepackage[numbers]{natbib}
\usepackage{units}
\usepackage{graphicx}
\usepackage{amsmath}
\usepackage{aecompl}
\usepackage{algorithm,algpseudocode}

\newenvironment{sciabstract}{%
\begin{quote} \bf}
{\end{quote}}

\newcounter{lastnote}

\title{A Detection Metric Designed for O'Connell Effect Eclipsing Binaries}

\author
{Kyle B. Johnston,$^{1, 5\ast}$ Rana Haber,$^{2}$ \\
Saida M. Caballero-Nieves, $^{1}$ Adrian M. Peter, $^{3}$ \\
V\'eronique Petit, $^{4}$ Matt Knote, $^{1}$\\
\\
\normalsize{$^{1}$Aerospace, Physics and Space Sciences Department, Florida Institute of Technology,}\\
\normalsize{150 West University Blvd., Melbourne FL., USA}\\
\normalsize{$^{2}$Mathematical Sciences Department, Florida Institute of Technology,}\\
\normalsize{150 West University Blvd., Melbourne FL., USA}\\
\normalsize{$^{3}$Computer Engineering and Sciences Department, Florida Institute of Technology,}\\
\normalsize{150 West University Blvd., Melbourne FL., USA}\\
\normalsize{$^{4}$Physics and Astronomy Department, University of Delaware,}\\
\normalsize{217 Sharp Lab, Newark, DE., USA}\\
\normalsize{$^{5}$Defense Group Melbourne, Perspecta Inc.,}\\
\normalsize{4849 N. Wickham Rd., Melbourne, FL., USA}\\
\\
\normalsize{$^\ast$To whom correspondence should be addressed; E-mail:  kyjohnst2000@my.fit.edu.}
}

\date{}

\begin{document} 


\baselineskip24pt


\maketitle


\begin{sciabstract}
We present the construction of a novel time-domain signature extraction
methodology and the development of a supporting supervised pattern
detection algorithm. We focus on the targeted identification
of eclipsing binaries that demonstrate a feature known as the O'Connell
effect. Our proposed methodology maps stellar variable observations
to a new representation known as distribution fields (DFs). Given this
novel representation, we develop a metric learning technique directly
on the DF space that is capable of specifically identifying our stars of interest. The metric is tuned on a set of labeled eclipsing binary data from
the Kepler survey, targeting particular systems exhibiting the O'Connell
effect. The result is a conservative selection of 124 potential targets
of interest out of the Villanova Eclipsing Binary Catalog. Our framework
demonstrates favorable performance on Kepler eclipsing binary data,
taking a crucial step in preparing the way for large-scale data volumes
from next-generation telescopes such as LSST and SKA.
\end{sciabstract}


\section{Introduction}

With the rise of large-scale surveys, such as Kepler, the Transiting
Exoplanet Survey Satellite (TESS), the Kilodegree Extremely Little
Telescope (KELT), the Square Kilometre Array, and the Large Synoptic Survey
Telescope (LSST), a fundamental working
knowledge of statistical data analysis and data management to
reasonably process astronomical data is necessary. The ability
to mine these data sets for new and interesting astronomical information
opens a number of scientific windows that were once closed by poor
sampling, in terms of both number of stars (targets) and depth of
observations (number of samples).

This article focuses on the development of a novel, modular time-domain signature
extraction methodology
and its supporting supervised pattern detection algorithm for variable star detection. 
The design could apply to any number of variable star types that exhibit 
consistent periodicity (cyclostationary) in their flux; examples include most Cepheid-type
stars (RR Lyr, SX Phe, Gamma Dor, etc...) as well as other eclipsing binary types.
Nonperiodic variables would require a different feature space \citep{johnston2017variable},
but the underlying detection scheme could still be relevant. Herein we present the 
design's utility, by its targeting of eclipsing binaries that demonstrate a feature
known as the O'Connell effect.

We have selected O'Connell effect eclipsing binaries (OEEBs) to demonstrate initially our
detector design. We highlight OEEBs here because they compose a subclass of a specific type 
of variable star (eclipsing binaries). Subclass detection provides an extra layer of complexity 
for our detector to try to handle. We demonstrate our detector design on Kepler eclipsing
binary data from the Villanova catalog, allowing us to train 
and test against different subclasses in the same parent variable
class type. We train our detector design on Kepler eclipsing
binary data and apply the detector to a different survey---the Lincoln Near-Earth 
Asteroid Research asteroid survey \citep[LINEAR, ][]{stokes2000lincoln}---to demonstrate the
algorithm's ability to discriminate and detect our targeted subclass given 
not just the parent class but other classes as well.

Classifying variable stars relies on proper selection
of feature spaces of interest and a classification framework that
can support the linear separation of those features. Selected features should
quantify the telltale signature of the variability---the structure and information 
content. Prior studies to develop both features and classifiers include expert 
selected feature efforts 
\citep{Debosscher2009,Sesar2011,Richards2012,graham2013machine,armstrong2015k2,mahabal2017deep,hinners2018machine},
automated feature selection efforts \citep{mcwhirter2017classification,naul2018recurrent},
and unsupervised methods for feature extraction \citep{valenzuela2017unsupervised,modak2018unsupervised}.
The astroinformatics community-standard features include quantification
of statistics associated with the time-domain photometric data, Fourier
decomposition of the data, and color information in both the optical
and IR domains \citep{nun2015fats,miller2015machine}. The number of
individual features commonly used is upward of 60 and growing \citep{richards2011machine}
as the number of variable star types increases, and as a result of
further refinement of classification definitions \citep{kazarovets2017general}. We seek here to develop a novel feature space that captures the signature of 
interest for the targeted variable star type.  

The detection framework here maps time-domain stellar variable observations
to an alternate distribution field (DF) representation \citep{sevilla2012distribution}
and then develops a metric learning approach to identify OEEBs. Based on the
matrix-valued DF feature, we adopt a metric learning framework to
directly learn a distance metric \citep{bellet2015metric} on the space of DFs. We can then utilize the learned
metric as a measure of similarity to detect new
OEEBs based on their closeness to other OEEBs. We present our metric learning
approach as a competitive push--pull optimization, where
DFs corresponding to known OEEBs influence the learned metric to measure
them as being nearer in the DF space. Simultaneously, DFs corresponding
to non-OEEBs are pushed away and result in large measured distances
under the learned metric. 

This article is structured as follows. First, we review the targeted stellar variable type, discussing the
type signatures expected. Second, we review the 
data used in our training, testing, and discovery process as part of our demonstration of design. 
Next, we outline the novel proposed pipeline for OEEB detection; this review includes the feature space used, the designed detector/classifier, and the associated
implementation of an anomaly detection algorithm \citep{Chandola2009}. Then, we apply the algorithm, 
trained on the expertly 
selected/labeled Villanova Eclipsing Binary catalog OEEB targets, to the rest
of the catalog with the purpose of identifying new OEEB stars. We present the
results of the discovery process
using a mix of clustering and derived statistics. We apply the Villanova Eclipsing
Binary catalog trained classifier, without additional training,
to the LINEAR data set. We provide results of this cross-application, i.e., the
set of discovered OEEBs. For comparison, we detail two
competing approaches. We develop training and testing strategies for our metric 
learning framework, and finally, we conclude with a summary of our findings and 
directions for future research.

\section{Eclipsing Binaries with O’Connell Effect}
\begin{figure}
\begin{centering}
\includegraphics[scale=0.35]{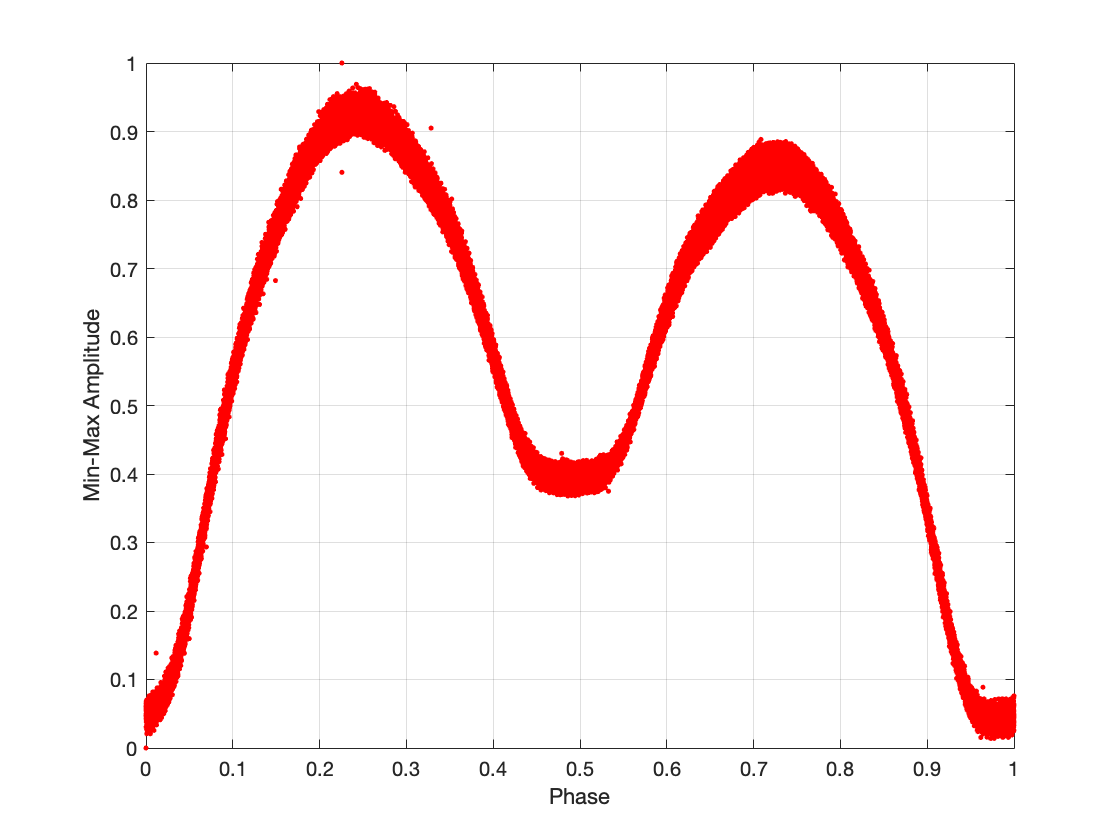}
\par\end{centering}
\caption{An example phased light curve of an eclipsing binary with the O'Connell effect (KIC: 10861842). The light curve has been phased such that the global minimum (cooler in front of hotter) is at lag 0 and the secondary
minimum (hotter in front of cooler) is at approximately lag 0.5. The
side-by-side binary orientations are at approximately 0.25 and
0.75. Note that the maxima, corresponding to the side-by-side orientations, have different values. \label{fig:ExampleOEEB}}
\end{figure}

The O'Connell effect \citep{o1951so} is defined for eclipsing binaries
as an asymmetry in the maxima of the phased light curve (see Figure \ref{fig:ExampleOEEB}). 
This maxima asymmetry is unexpected, as it suggests an orientation dependency
in the brightness of the system. Similarly, the consistency of the
asymmetric over many orbits is also surprising, as it suggests that the
maxima asymmetry has a dependence on the rotation of the binary system. The cause of the
O’Connell effect is not fully understood and additional data and modeling are necessary 
for further investigation \citep{McCartney1999}. Our focus in this work is in the 
application of an automated detector to OEEBs to identify systems of interest for future work.

\subsection{Signatures and Theories}

Several theories propose to explain the effect, including starspots, gas stream impact, and circumstellar
matter \citep{McCartney1999}. The work by \citep{wilsey2009revisiting}
outlines each of these theories and demonstrates how the observed
effects are generated by the underlying physics.

\begin{itemize}
\item Starspots result from chromospheric activity, causing a consistent 
decrease in brightness of the star when viewed as a point source. While magnetic surface activity will cause both flares (brightening) and spots (darkening), flares tend to be transient, whereas spots tend to have longer-term effects on the observed binary flux. Thus, between the two, starspots are the favored hypothesis for causing long-term consistent asymmetry; often binary simulations (such as the Wilson--Devinney code) can be used to model O'Connell effect  binaries via including an often large starspot \citep{Zboril2006}. 
\item Gas stream impact results from matter transferring between stars (smaller
to larger) through the L1 point and onto a specific position on the
larger star, resulting in a consistent brightening on the leading/trailing
side of the secondary/primary. 
\item The circumstellar matter theory proposes to describe the increase
in brightness via free-falling matter being swept up, resulting in
energy loss and heating, again causing an increase in amplitude. Alternatively,
circumstellar matter in orbit could result in attenuation, i.e.,
the difference in maximum magnitude of the phased light curve results
from dimming and not brightening.
\end{itemize}

In the study \cite{McCartney1999}, the authors limited the sample to only
six star systems: GSC 03751-00178, V573 Lyrae, V1038 Herculis, ZZ
Pegasus, V1901 Cygni, and UV Monocerotis. Researchers have used standard eclipsing binary
simulations \citep{wilson1971realization} to demonstrate
the proposed explanations for each light curve instance and estimate
the parameters associated with the physics of the system. \citep{wilsey2009revisiting}
noted other cases of the O'Connell effect in binaries, which have since been described
physically; in some cases, the effect varied over time, whereas in other cases, the effect was consistent over years of observation and over many orbits. The effect has been found in both overcontact, semidetached, and near-contact systems. 

While one of the key visual differentiators of the O'Connell effect is $\Delta m_{\rm max}$, this heuristic feature alone could not be used as a general mean for detection, as the targets trained on or applied to are not guaranteed to be (a) eclipsing binaries and (b) periodic. One of the goals we are attempting to highlight is the transformation of expert qualitative target selection into  quantitative machine learning methods.

\subsection{Characterization of OEEB}

We develop a detection methodology for a specific target of interest---OEEB---defined
as an eclipsing binary where the light curve (LC) maxima are consistently at
different amplitudes over the span of observation. Beyond differences
in maxima, and a number of published examples, little is defined as
a requirement for identifying the O'Connell
effect \citep{wilsey2009revisiting,knote:inpress-b}.

\cite{McCartney1999} provide some basic indicators/measurements of interest in 
relation to OEEB binaries: the O'Connell effect ratio
(OER), the difference in maximum amplitudes ($\Delta m$), the difference
in minimum amplitudes, and the light curve asymmetry (LCA). The
metrics are based on the smoothed phased light curves. The OER is calculated as Equation \ref{eq:OER}:
\begin{equation}
{\rm OER}=\frac{\sum_{i=1}^{\nicefrac{n}{2}}\left(I_{i}-I_{1}\right)}{\sum_{i=\nicefrac{n}{2}+1}^{n}\left(I_{i}-I_{1}\right)},\label{eq:OER}
\end{equation}
where the min-max amplitude (i.e. normalized flux) measurements for each star are grouped
into phase bins ($n=500$), where the mean amplitude in each bin is $I_{i}$.
An $OER>1$ corresponds to the front half of the light curve having
more total flux; note that for the procedure we present here, $I_{1}=0$.
The difference in max amplitude is calculated as Equation \ref{eq:deltamax}:
\begin{equation}
\Delta m=\max_{t<0.5}\left(f(t)_N\right)-\max_{t\geq0.5}\left(f(t)_N\right),\label{eq:deltamax}
\end{equation}
where we have estimated the maximum in each half of the phased light
curve. The LCA is calculated as Equation \ref{eq:lca}:
\begin{equation}
{\rm LCA}=\sqrt{\sum_{i=1}^{\nicefrac{n}{2}}\frac{\left(I_{i}-I_{\left(n+1-i\right)}\right)^{2}}{I_{i}^{2}}}.\label{eq:lca}
\end{equation}

As opposed to the measurement of OER, LCA measures the deviance
from symmetry of the two peaks. Defining descriptive metrics or 
functional relationships (i.e., bounds of distribution)
requires a larger sample than is presently available. An increased number of 
identified targets of interest is required to provide the sample size needed for a complete statistical description of the O'Connell effect. The quantification of these functional statistics allows for the improved understanding of not just the standard definition of the targeted variable star but also the population distribution as a whole. These estimates allow for empirical statements to be made regarding the differences in light curve shapes among the variable star types investigated. The determination of an empirically observed distribution, however, requires a significant sample to generate meaningful descriptive statistics for the various metrics. 

In this effort, we highlight the functional shape of the phased light curve 
as our defining feature of OEEB stars. The prior metrics identified are selected 
or reduced measures of this functional shape. We propose here that, as opposed to
training a detector on the preceding indicators, we use the functional shape 
of the phased light curve by way of the distribution field to construct our automated system. 

\section{Variable Star Data}

As a demonstration of design, we apply the proposed algorithm to
a set of predefined, expertly labeled eclipsing binary light curves. We focus on two
surveys of interest: first, the Kepler Villanova Eclipsing Binary catalog, from which
we derive our initial training data as well as our initial discovery (unlabeled) 
data, and second, the  Lincoln Near-Earth Asteroid 
Research, which we treat as unlabeled data.

\subsection{Kepler Villanova Eclipsing Catalog}

Leveraging the Kepler pipeline already in
place, and using the data from the Villanova Eclipsing Binary catalog
\citep{kirk2016kepler}, this study focuses on a set of predetermined
eclipsing binaries identified from the Kepler catalog. From this catalog,
we developed an initial, expertly derived, labeled data set of proposed
targets ``of interest'' identified as OEEB. Likewise, we generated a set of targets
identified as ``not of interest'' based on our expert definitions,
i.e., intuitive inference. 

We have labeled our two populations ``of interest'' and ``not of interest'' to represent those targets in the initial training set that a user has found to either be interesting to their research (identified OEEBs) or otherwise. The labeling ``of interest'' here is specifically used, as we are dependent on the expert selections.

Using the Eclipsing Binary catalog \citep{kirk2016kepler}, we identified a set of 30 targets ``of interest''
and 121 targets of ``not of interest'' via expert analysis---by-eye 
selection based on researchers' interests. Specific target identification 
is listed in a supplementary digital file at the project repository.\footnote{https://github.com/kjohnston82/OCDetector/supplement/KeplerTraining.xlsx}
We use this set of 151 light curves for training and testing.

\subsubsection{Light Curve/Feature Space\label{subsec:Signal-Conditioning-and}}

Prior to feature space processing, the raw observed photometric time
domain data are conditioned and processed. Operations include long-term trend removal, artifact removal, initial light curve phasing,
and initial eclipsing binary identification; we performed these actions
prior to the effort demonstrated here, by the Eclipsing Binary catalog
(our work uses all 2875 long-cadence light curves available as of
the date of publication as training/testing data, or as unlabeled data to search for new OEEBs). The functional shape of the phased light
curve is selected as the feature to be used in the machine learning
process, i.e., detection of targets of interest. While the data have
been conditioned already by the Kepler pipeline, added steps are taken
to allow for similarity estimation between phased curves. Friedman's
SUPERSMOOTHER algorithm \citep{friedman1984variable, VanderPlas2015} is used to generate
a smooth 1-D functional curve from the phased light curve data. The
smoothed curves are transformed via min-max scaling \ref{eq:1}:
\begin{equation}
f(\phi)_N=\frac{f(\phi)-\min(f(\phi))}{\max\left(f(\phi)\right)-\min(f(\phi))},\label{eq:1}
\end{equation}
where $f(\phi)$ is the smoothed phased light curve, $f$ is the amplitude
from the database source, $\phi$ is the phase where $\phi\in[0,1]$,
and $f(\phi)_N$ is the min-max scaled amplitude (i.e. normalized flux). Note that we will use the 
terms $f(\phi)_N$ and min-max amplitude interchangeably throughout this article. We use the minimum of the
smoothed phased light curve as a registration marker, and both
the smoothed and unsmoothed light curves are aligned such that lag/phase
zero corresponds to minimum amplitude (eclipse minima; see \citep{McCartney1999}).

\subsubsection{Training/Testing Data}
The labeled training data are provided as part of the supplementary digital 
project repository. We include the SOI and NON-SOI Kepler identifiers here (KIC).

\begin{table}[H]
\caption{Collection of KIC of Interest (30 Total)\label{SOI_OEEB}}
\centering{}%
\begin{tabular}{cccccccc}
\hline 
10123627 &	11924311&	5123176 & 8696327 &	11410485 &	7696778 & 7516345 &	9654476 \\
10815379 &	2449084	&	5282464	& 8822555 &	7259917  &	6223646 & 4350454 &	9777987 \\
10861842 &	2858322	&	5283839	& 9164694 &	7433513  &	9717924 & 5820209 & 7584739 	\\
11127048 & 4241946 & 5357682 & 9290838 & 8394040 &	7199183	\\
\hline 
\end{tabular}
\end{table}

\begin{table}[H]
\caption{Collection of KIC Not of Interest (121 Total)\label{NONSOI_OEEB}}
\centering{}%
\begin{tabular}{cccccccc}
\hline 
10007533 &	10544976 &	11404698 & 5560831  & 7119757	 &	5685072 &	7335517	& 5881838  \\ 
10024144 &	10711646 &	11442348 &	12470530 &	3954227 & 10084115 &	10736223 &	11444780 \\ 
10095469 &	10794878 &	11652545 &	2570289	&	4037163 & 10216186 &	10802917 &	12004834 \\
10253421 &	10880490 &	12108333 &	3127873	&	4168013 & 10257903 &	10920314 &	12109845  \\
10275747 &	11076176 &	12157987 &	3344427	&	4544587 & 10383620 &	11230837 &	12216706  \\
10485137 &	11395117 &	12218858 &	3730067 &	4651526 & 9007918 &	8196180	 &	7367833	  \\
9151972 &	8248812	 &	7376500	 &	6191574	&	4672934 & 9179806 &	8285349	 &	7506446	 \\
9205993 &	8294484	 &	7518816	 &	6283224 &	4999357 & 9366988 &	8298344	 &	7671594  \\
9394601 &	8314879	 &	7707742  &	6387887	&	5307780 & 9532219 &	8481574	 &	7879399 \\
9639491 &	8608490	 &	7950964	 &	6431545	&	5535061 & 9700154 &	8690104	 &	8074045  \\
9713664 &	8758161	 &	8087799	 &	6633929 &	5606644 & 9715925 &	8804824	 &	8097553	  \\
9784230	&	8846978	 &	8155368	 &	7284688 &	5785551 & 9837083 &	8949316 &	8166095	  \\
9935311 &	8957887  &	8182360	 &	7339345 &	5956776 & 9953894 &	3339563	 &	4474193	 \\
12400729 &	3832382  & 12553806  &	4036687 &	2996347	& 4077442 &	3557421	 &	4554004\\
6024572	 &	4660997  &	6213131  &	4937217 &	6370361 & 5296877 &	6390205	 &	5374999\\
6467389\\
\hline
\end{tabular}
\end{table}

\begin{figure}
\begin{centering}
\includegraphics[scale=0.35]{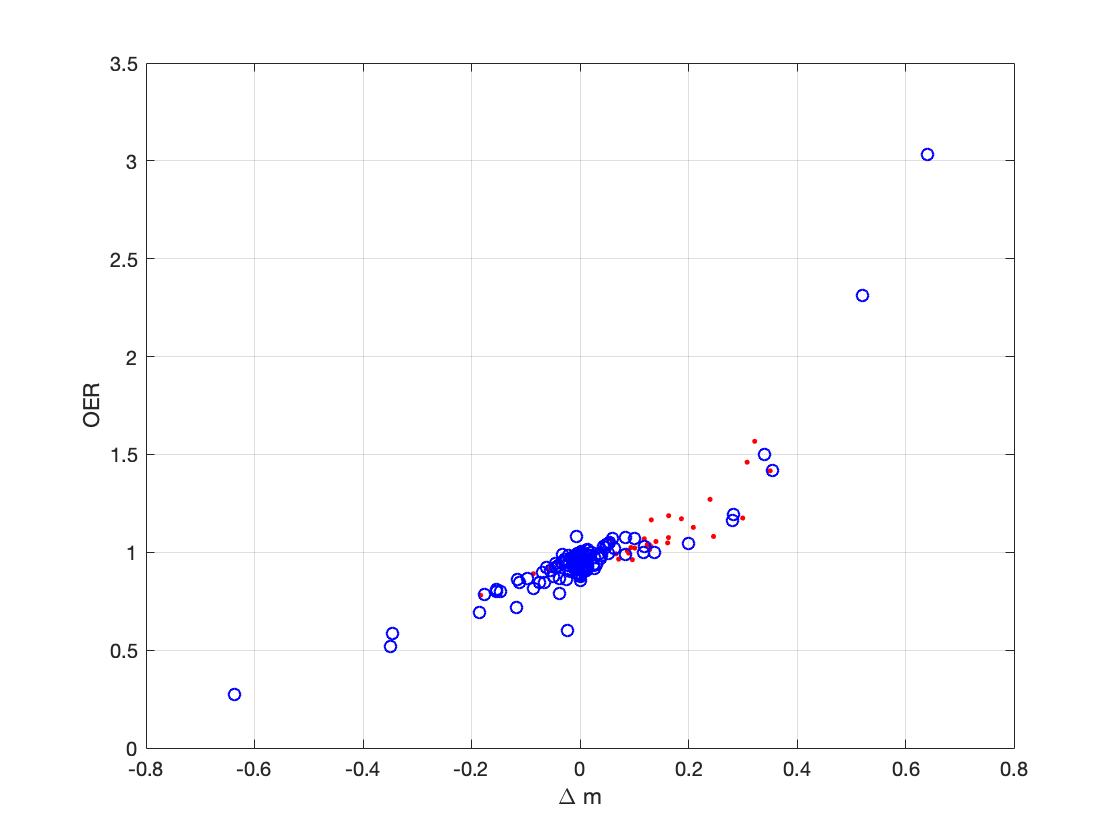}
\par\end{centering}
\caption{The features $\Delta m$ and $\rm OER$ for the training data; red dots are the targets are ``Of Interest'', and blue circles are ``Not of Interest''. \label{fig:ExampleTraining}}
\end{figure}

Additionally, we plot the total training and testing dataset in the $\Delta m$ and $\rm OER$ feature space, to demonstrate the separability of our classes (``Of Interest'' vs. ``Not of Interest''), see figure \ref{fig:ExampleTraining}. The values presented here were generated based on phased, unsmoothed, data.

As is apparent from figure \ref{fig:ExampleTraining}, the data does not exactly separate based on either of the select heuristic measures into the selected categories. For a baseline estimate of performance, we use a simple 1-NN classification algorithm with our selected two heuristic $\Delta m$ and $\rm OER$ values, using a randomized 50/50 split of our initial Kepler training data (see section \ref{subsec:Training,-Cross-Validation-and}), resulting in a misclassification rate of $23\%$. This error rate drops to $14\%$ if we use k-NN (with a $k = 5$); larger values of k, consistently decreased performance of the classifier further.

The training data is based on expert requests, with the explicit request that the detector find new observations that are similar to those ``of interest'' and dissimilar to those identified as ``not of interest''. These targets were to have a   $\left|\Delta m\right|$ greater than some threshold, was not to have multiple periods, and was to have a consistent structure in the phased domain. Our objective was to construct a procedure that could find other light curves that fit these user constraints.

\subsection{Data Untrained}

The 2,000+ eclipsing binaries left in the Kepler Eclipsing Binary catalog are left 
as unlabeled targets. We use our described detector to ``discover'' targets of interest,
i.e., OEEB. The full set of Kepler data is accessible via the Villanova Eclipsing
Binary website (http://keplerebs.villanova.edu/).

For analyzing the proposed algorithm design, the LINEAR data set is also 
leveraged as an unknown ``unlabeled'' data set ripe for OEEB discovery \citep{Sesar2011,Palaversa2013}.
From the starting sample of 7,194 LINEAR variables, we used a clean sample
of 6,146 time series data sets for detection. Stellar class type is
limited further to the top five most populous classes---RR Lyr (ab), RR Lyr (c), Delta
Scuti / SX Phe, Contact Binaries, and Algol-Like Stars with two minima---resulting in a set of 5,086 observations.

Unlike the Kepler Eclipsing Binary catalog, the LINEAR data set contains
targets other than (but does include) eclipsing binaries; the data set we used \citep{johnston2017variable} 
includes Algols (287), Contact Binaries (1805), Delta Scuti (68), and RR Lyr (ab-2189, c-737). The light curves are much more poorly sampled; this uncertainty
in the functional shape results from lower SNR (ground survey) and
poor sampling. The distribution of stellar
classes is presented in Table \ref{TableDistribution_LINEAR}.

\begin{table}[H]
\caption{Distribution of LINEAR Data across Classes\label{TableDistribution_LINEAR}}
\centering{}%
\begin{tabular*}{\textwidth}{@{\extracolsep\fill}lcc}
\hline 
Type & Count & Percentage\\
\hline 
Algol & 287 & 5.6\\
Contact Binary & 1805 & 35.6\\
Delta Scuti & 68 & 1.3\\
RRab & 2189 & 43.0\\
RRc & 737 & 14.5\\
\hline 
\end{tabular*}
\end{table}

The full data sets used at the time
of this publication from the Kepler and LINEAR surveys are available from the associated public 
repository.\footnote{github.com/kjohnston82/OCDetector}

\section{PMML Classification Algorithm}

Relying on previous designs in astroinformatics to develop a supervised
detection algorithm \citep{johnston2017generation}, we propose a design
that tailors the requirements specifically toward detecting
OEEB-type variable stars. 

\subsection{Prior Research}

Many prior studies on time-domain variable star classification \citep{Debosscher2009,barclay2011stellar,
Blomme2011,dubath2011random,Pichara2012,Pichara2013,graham2013machine,Angeloni2014,Masci2014}
rely on periodicity domain feature space reductions. \cite{Debosscher2009}
and \cite{templeton2004time} review a number of feature spaces and
a number of efforts to reduce the time-domain data, most of which
implement Fourier techniques, primarily the Lomb--Scargle (L-S) method
\citep{lomb1976least,scargle1982studies}, to estimate the primary
periodicity \citep{eyer2005automated,deb2009light,Richards2012,Park2013,ngeow2013preliminary}.

The studies on classification of time-domain variable stars often further reduce the folded time-domain data into features that provide maximal-linear separability of classes. 
These efforts include expert selected feature efforts \citep{Debosscher2009, Sesar2011, Richards2012,
graham2013machine, armstrong2015k2, mahabal2017deep, hinners2018machine},
automated feature selection efforts \citep{mcwhirter2017classification,naul2018recurrent},
and unsupervised methods for feature extraction \citep{valenzuela2017unsupervised,modak2018unsupervised}.
The astroinformatics community-standard features include quantification
of statistics associated with the time-domain photometric data, Fourier
decomposition of the data, and color information in both the optical
and IR domains \citep{nun2015fats,miller2015machine}. The number of
individual features commonly used is upward of 60 and growing \citep{richards2011machine}
as the number of variable star types increases and as a result of
further refinement of classification definitions \citep{kazarovets2017general}. 
Curiously, aside from efforts to construct a classification algorithm
from the time-domain data directly \citep{mcwhirter2017classification},
few efforts in astroinformatics have looked at features beyond
those described here---mostly Fourier domain transformations or time
domain statistics. Considering the depth of possibility
for time-domain transformations \citep{Fu2011,Grabocka2012,cassisi2012similarity,fulcher2013highly},
it is surprising that the community has focused on just a few transforms.  Additionally, there has been recent work in exo-planet detection using whole phased waveform data (smoothed), in combination with neural network classification \citep{pearson2017searching}, and with local linear embedding \citep{thompson2015machine}. Similar to the design proposed here, these methods use the classifier to optimize the feature space (the phased waveform) for the purposes of detection. 

Here we propose an implementation that simplifies the traditional design: 
limiting ourselves to a
one versus all approach \citep{johnston2017generation} targeting a variable type 
of interest; limiting ourselves to a singular feature space---the distribution field
of the phased light curve---based on \cite{Helfer2015} as a representation of the 
functional shape; and introducing a classification/detection scheme that is based on 
similarity with transparent results \citep{bellet2015metric} that can be further 
extended, allowing for the inclusion of an anomaly detection algorithm.

\subsection{Distribution Field\label{subsec:distrfield}}

As stated, this analysis focuses on detecting OEEB systems
based on their light curve shape. The OEEB signature has a cyclostationary
signal, a functional shape that repeats with a consistent frequency.
The signature can be isolated using a process of period finding, folding,
and phasing \citep{graham2013comparison};  the Villanova catalog provides the estimated ``best
period.'' The proposed feature
space transformation will focus on the quantification or representation
of this phased functional shape. This particular implementation design
makes the most intuitive sense, as visual inspection of the phased
light curve is the way experts identify these unique sources. 

As discussed, prior research on time-domain data identification has
varied between generating machine-learned features \citep{gagniuc2017markov},
implementing generic features \citep{Masci2014,Palaversa2013,Richards2012,Debosscher2009},
and looking at shape- or functional-based features \citep{haber2015discriminative,johnston2017variable,Park2013}.
This analysis will leverage the distribution field transform to generate
a feature space that can be operated on; a distribution field (DF)
is defined as \citep{Helfer2015,sevilla2012distribution} Equation
\ref{eq:2}:
\begin{equation}
DF_{ij}=\frac{\sum_k^N\left[y_{j}<f\left(x_{i}\leq\phi_k\leq x_{i+1}\right)_N<y_{j-1}\right]}{\sum_k^N\left[y_{j}<f\left(\phi_k\right)_N<y_{j-1}\right]},\label{eq:2}
\end{equation}
where N is the number of samples in the phased data, and  $\left[\:\right]$ is the Iverson bracket \citep{iverson1962programming},
given as
\begin{equation}
[P]=\left\{ \begin{array}{cc}
1 & P={\rm true}\\
0 & {\rm otherwise,}
\end{array}\right.\label{eq:iverson}
\end{equation}
and $y_{j}$ and $x_{i}$ are the corresponding normalized amplitude
and phase bins, respectively, where $x_i = {0, 1/n_x, 2/n_x, \dots, 1}$,
$y_i = {0, 1/n_y, 2/n_y, \dots, 1}$, $n_x$ is the number of time
bins, and $n_y$ is the number of amplitude
bins. The result is a right stochastic matrix, i.e., the rows sum to 1. 
Bin number, $n_x$ and $n_y$, is optimized by cross-validation as
part of the classification training process. Smoothed phased data
---generated from SUPERSMOOTHER---are provided to the DF algorithm.

\begin{figure}
\begin{centering}
\includegraphics[scale=0.35]{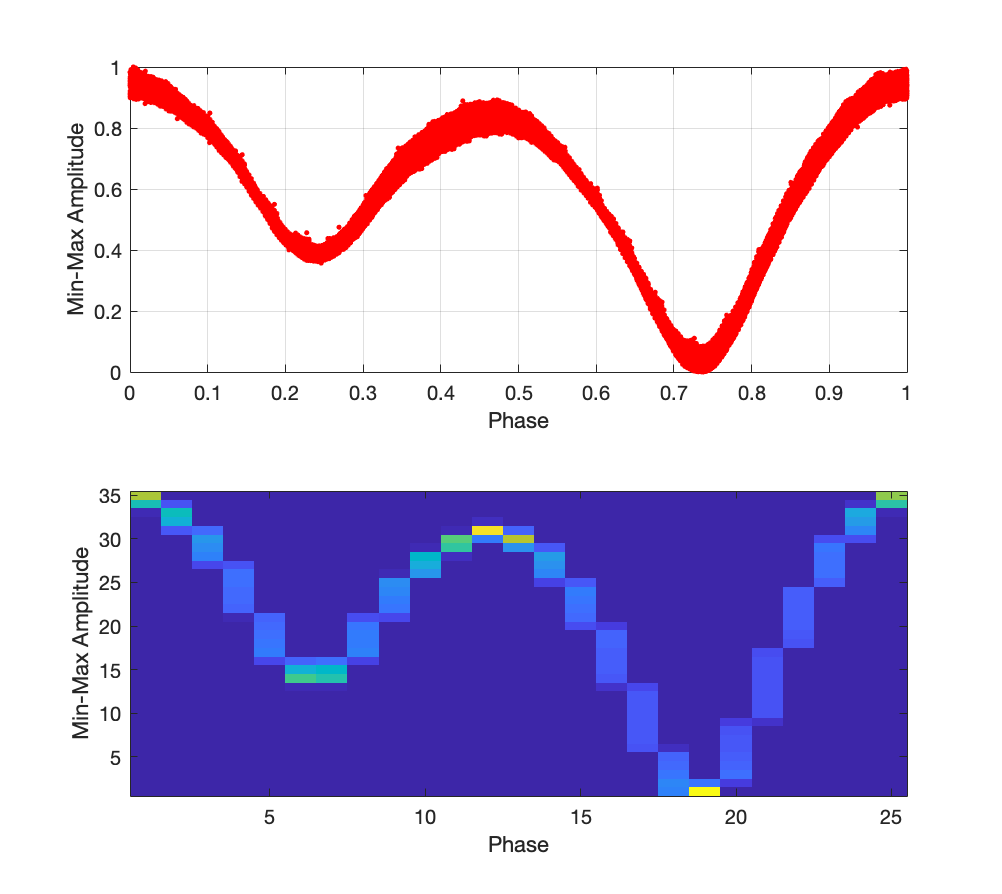}
\par\end{centering}
\caption{An example phased light curve (top) and the transformed distribution
field (bottom) of an Eclipsing Binary with the O'Connell
effect (KIC: 7516345). \label{fig:ExampleDF}}
\end{figure}

We found this implementation to produce a more consistent classification
process. We found that the min-max scaling normalization---normalized flux---if applied by itself without smoothing, when
outliers are present can produce final patterns that focus more on
the outlier than the general functionality of the light curve. Likewise, 
we found that using the unsmoothed data in the DF algorithm resulted in a 
classification that was too dependent on the scatter of the phased light curve.
Although at first glance, that would not appear to be an issue, this implementation 
resulted in light curve resolution having a large impact on the classification 
performance---in fact, a higher impact than the shape itself. An example of this transformation is given in Figure \ref{fig:ExampleDF}.

Though the DF exhibits properties that a detection
algorithm can use to identify specific variable stars of interest, it alone
is not sufficient for our ultimate goal of automated detection. Rather
than vectorizing the DF matrix and treating it as a feature vector
for standard classification techniques, we treat the DF as the matrix-valued
feature that it is \citep{Helfer2015}. This allows for the retention
of row and column dependence information that would normally be lost
in the vectorization process \citep{ding2018matrix}.

\subsection{Metric Learning\label{subsec:Push-Pull-Matrix-Metric}}

At its core, the proposed detector is based on the definition of similarity
and, more formally, a definition of distance. Consider the example triplet
``$x$ is more similar to $y$ than to $z$,'' i.e., the distance
between $x$ and $y$ in the feature space of interest is smaller
than the distance between $x$ and $z$. The field of metric learning
focuses on defining this distance in a given feature space to
optimize a given goal, most commonly the reduction of error rate associated
with the classification process. Given the selected feature space of DF matrices, the distance between two matrices $X$ and $Y$ \citep{bellet2015metric, Helfer2015}
is defined as Equation \ref{eq:3}:
\begin{equation}
d(X,Y)=\left\Vert X-Y\right\Vert _{M}^{2}=tr\left\{ \left(X-Y\right)^{T}M\left(X-Y\right)\right\}. \label{eq:3}
\end{equation}
$M$ is the metric that we will be attempting to optimize, where $M\succeq0$ 
(positive semi-definite). The PMML procedure outlined
in \citep{Helfer2015} is similar to the metric learning methodology
LMNN \citep{weinberger2006distance}, save for its implementation on matrix-variate 
data as opposed to vector-variate data. We summarize it here. The 
developed objective function is given in Equation \ref{eq:4}:
\begin{multline}
E=\frac{1-\lambda}{N_{c}-1}\sum_{i,j}\left\Vert {\rm DF}_{c}^{i}-{\rm DF}_{c}^{j}\right\Vert _{M}^{2}\\
-\frac{\lambda}{N-N_{c}}\sum_{i,k}\left\Vert {\rm DF}_{c}^{i}-{\rm DF}_{c}^{k}\right\Vert _{M}^{2}+\frac{\gamma}{2}\left\Vert M\right\Vert _{F}^{2},\label{eq:4}
\end{multline}
where $N_c$ is the number of training data in class $c$; $\lambda$ and $\gamma$ are 
variables to control the importance of push versus pull and regularization, respectively. \cite{bellet2015metric} define the triplet $\left\{ {\rm DF}_{c}^{i},{\rm DF}_{c}^{j},{\rm DF}_{c}^{k}\right\} $ as the
relationship between similar and dissimilar observations, i.e., ${\rm DF}_{c}^{i}$ is similar to ${\rm DF}_{c}^{j}$ and dissimilar to ${\rm DF}_{c}^{k}$.Note, the summation over $i$ and $j$ is the summation over similar observations in the training data, and the summation $i$ and $k$ is the summation over dissimilar observations.

There are three basic components: a pull term, which is small
when the distance between similar observations is small; a push term,
which is small when the distance between dissimilar observations is
larger; and a regularization term, which is small when the Frobenius
norm ($\left\Vert M\right\Vert _{F}^{2} = \sqrt{Tr(MM^H)}$) of $M$ is small. Thus the algorithm attempts to bring similar distribution fields closer together, while pushing dissimilar ones farther apart, while attempting to minimize the complexity of the metric $M$. The regularizer on the metric $M$ guards against overfitting
and consequently enhances the algorithm's ability to generalize, i.e.,
allow for operations across data sets.This regularization strategy
is similar to popular regression techniques like lasso and ridge \citep{Hastie2009}. 

Additional parameters $\lambda$ and $\gamma$ weight the importance
of the push--pull terms and metric regularizer, respectively. These
free parameters are typically tuned via standard cross-validation
techniques on the training data. The objective function represented
by Equation \ref{eq:4} is quadratic in the unknown metric $M$; hence
it is possible to obtain the following closed-form solution to the minimization of Equation \ref{eq:4} as:
\begin{multline}
M=\frac{\lambda}{\gamma\left(N-N_{c}\right)}\sum_{i,k}\left({\rm DF}_{c}^{i}-{\rm DF}_{c}^{k}\right)\left({\rm DF}_{c}^{i}-{\rm DF}_{c}^{k}\right)^{T}\\
-\frac{1-\lambda}{\gamma\left(N_{c}-1\right)}\sum_{i,j}\left({\rm DF}_{c}^{i}-{\rm DF}_{c}^{j}\right)\left({\rm DF}_{c}^{i}-{\rm DF}_{c}^{j}\right)^{T}.\label{eq:5}
\end{multline}

Equation \ref{eq:5} does not guarantee that $M$ is
positive semi-definite (PSD). To ensure this property, we can apply
the following straightforward projection step after calculating $M$ to ensure the requirement of $M\succeq0$:
\begin{enumerate}
\item perform eigen decomposition: $M=U^{T}\Lambda U$;
\item generate $\Lambda_{+}=\max\left(0,\Lambda\right)$, i.e., select positive
eigenvalues;
\item reconstruct the metric $M$: $M=U^{T}\Lambda_{+}U$. 
\end{enumerate}

If $M$ is not PSD, then the distance axioms are not held up, and therefore our similarity that we are using $d(X,Y)=\left\Vert X-Y\right\Vert _{M}^{2}$ would not be a true distance, specifically if M is not symmetric then $d(x_i, x_j) \neq d(x_j, x_i)$. This projected metric is used in the classification algorithm. The
metric learned from this push--pull methodology is used in conjunction
with a standard k-nearest neighbor (k-NN) classifier. 

\subsection{k-NN Classifier}

The traditional k-NN algorithm is a nonparametric classification
method; it uses a voting scheme based on an initial training set to
determine the estimated label \citep{altman1992introduction}. For a given 
new observation, the $L_{2}$ Euclidean distance is found between the new observation and all points
in the training set. The distances are sorted, and the $k$ closest
training sample labels are used to determine the new observed sample
estimated label (majority rule). Cross-validation is used to find
an optimal $k$ value, where $k$ is any integer greater than zero. 

The k-NN algorithm estimates a classification label based on the closest samples
provided in training. For our implementation, the distance between a new pattern
${\rm DF}^{i}$ and each pattern in the training set is found, using the optimized metric instead of the standard identity metric that would have been used in $L_{2}$ Euclidean distance. The new pattern
is classified depending on the majority of the closest $k$ class
labels. The distance between patterns is in Equation \ref{eq:3},
using the learned metric $M$.

\section{Results of the Classification}

The new OEEB systems discovered by the method of automated detection proposed here
can be used to further investigate their frequency of occurrence, provide constraints
on existing light curve models, and provide parameters to look for these systems
in future large-scale variability surveys like LSST.

\subsection{Training on Kepler Data \label{subsec:Training,-Cross-Validation-and}}

Optimized feature dimensions and parameters used in the classification process were estimated via five-fold cross-validation\citep{Duda2012}. Our procedure is as follows: the original set of 151 was split into two groups, 76 for training and 75 for testing. The training dataset is partitioned into 5 groups of equal sizes (14), with no replication. To evaluate the optimal DF dimensions we loop over a range of both $x$ and $y$ resolutions. We also include a loop for number of $k$-Neighbors. Within these loops we evaluate the performance of three classifiers that have limited input parameters to train using our partitioned data. This cycling over the partitioned data is the 5-fold cross-validation, within this loop we cycle over each partition, leaving it out, using the other four for training, and comparing the classifier trained on the four against the one left out.
 
The resulting error average over the five cycles is used as the performance estimate for the selection of $x$/$y$/$k$ resolution. After the resulting analysis, we have three error estimates, per $x$/$y$/$k$ resolution pairing. We select the ``optimal'' $x$/$y$ resolution pairing based on a minimization of the PMML classifier, for all classifiers trained (the optimal resolutions for the other classifier methods resulted in roughly the same resolution at the PMML one); $k$ was optimized per classifier. These optimal resolutions are then used to generate the DF features from all of the training data. This training data is then used to train the classifier selected, and those trained classifiers are then applied to the testing data that was originally set aside. The resulting misclassification rates are provided in the Table \ref{tab:PerformanceEstimates-1}.

We make the following general notes/caveats about the process used: 

\begin{itemize}
\item The selection of partitions is a random process (there is a random selection algorithm in the partitioning algorithm). Therefore the resulting errors produced as part of the cross-validation process aren't guaranteed to be the same on subsequent runs (i.e. different partitions selection = different error rates).
\item  Likewise, one of the alternative classification methodologies requires the use of clustering, so similar to (3) there is some inherent randomness associated with the process that may result in different results from run to run.
\end{itemize}

The minimization of misclassification rate is used to optimize floating
parameters in the design, such as the number of $x$-bins, the number
of $y$-bins, and $k$-values (i.e. k number of neighbors). The cross-validation process tested $n_x$ and $n_y$ values over values 20--40 in steps of 5. Some parameters are more sensitive
than others; often this insensitivity is related to the loss function
or the feature space, or the data themselves. For example, the $\gamma$ and $\lambda$ values weakly affected the optimization, while the bin sizes and $k$-values had a stronger effect ($\gamma$ and $\lambda$ values tested both spanned the range $0.0$ to $1.0$, $k$-Values were tested for odd values between $1$ and $13$, larger values of $k$ beyond those reported consistently decreased performance of the classifier).  

The set of optimized parameters is given as $\gamma=$1.0,
$\lambda=0.75$, $n_x = 25$, $n_y = 35$, and $k=3$. Given the optimization of these floating variables in all three algorithms,
the testing data are then applied to the optimal designs (testing results provided in Section \ref{subsec:results}).

\subsection{Application on Unlabeled Kepler Data\label{subsec:Application-to-Unlabeled}}

The algorithm is applied to the Villanova Eclipsing Binary catalog
entries that were not identified as either ``Of Interest'' or ``Not
of Interest,'' i.e., unlabeled for the purposes of our targeted goal. The trained and tested 
data sets are combined into a single training set for application; the primary method (push--pull
metric classification) is used to optimize a metric based on the optimal
parameters found during cross-validation and to apply the system to
the entire Villanova Eclipsing Binary data (2875 curves).

\subsubsection{Design Considerations}

On the basis of the results demonstrated in \citep{johnston2017generation}, the algorithm
additionally conditions the detection process based on a maximal distance
allowed between a new unlabeled point and the training data set in
the feature space of interest.

This anomaly detection algorithm is based on the optimized metric; a maximum distance 
between data points is based on the training data set, and we use a fraction (0.75) of
that maximum distance as a limit to determine ``known'' versus ``unknown.'' The 
value of the fraction was initially determined via trial and error, based on our 
experiences with the data set and the goal of minimizing false alarms 
(which were visually apparent). This further restricts the algorithm to classifying those 
targets that exist only in ``known space.'' The k-NN algorithm generates a distance
dependent on the optimized metric; by restricting the distances allowed, we can leverage
the algorithm to generate the equivalent of an anomaly detection algorithm.

The resulting paired algorithm (detector + distance limit) will produce estimates 
of ``interesting'' versus ``not interesting,'' given new---unlabeled---data. Our
algorithm currently will not produce confidence  estimates associated with the label.
Confidence associated with detection can be a touchy subject, both for the scientists 
developing the tools and for the scientists using them. Here we have
focused on implementing a k-NN algorithm with optimized metric (i.e., metric learning);
posterior probabilities of classification can be estimated based on
k-NN output \citep{Duda2012} and can be found as $(k_c/(n*{\rm volume}))$; linking these posterior
probability estimates to ``how confident am I that this is what I think this is'' is not
often the best choice of description. 
 
Confidence in our detections will be a function of
the original PPML classification algorithm performance, the training set used and the confidence 
in the labeling process, and the anomaly detection algorithm we implemented. Even $(k_c/(n*{\rm volume}))$
would not be a completely accurate description in our scenario. Some researchers \citep{dalitz2009reject} have worked on linking ``confidence'' in k-NN classifiers with distance 
between the points. Our introduction of an anomaly detection algorithm into the design thus
allows a developer/user the ability to limit the false alarm rate by introducing a maximum
acceptable distance thus allowing some control in the confidence of the result; see \citep{johnston2017generation} for more information.

\subsubsection{Results\label{subsec:results}}

\begin{figure}
\begin{centering}
\includegraphics[scale=0.35]{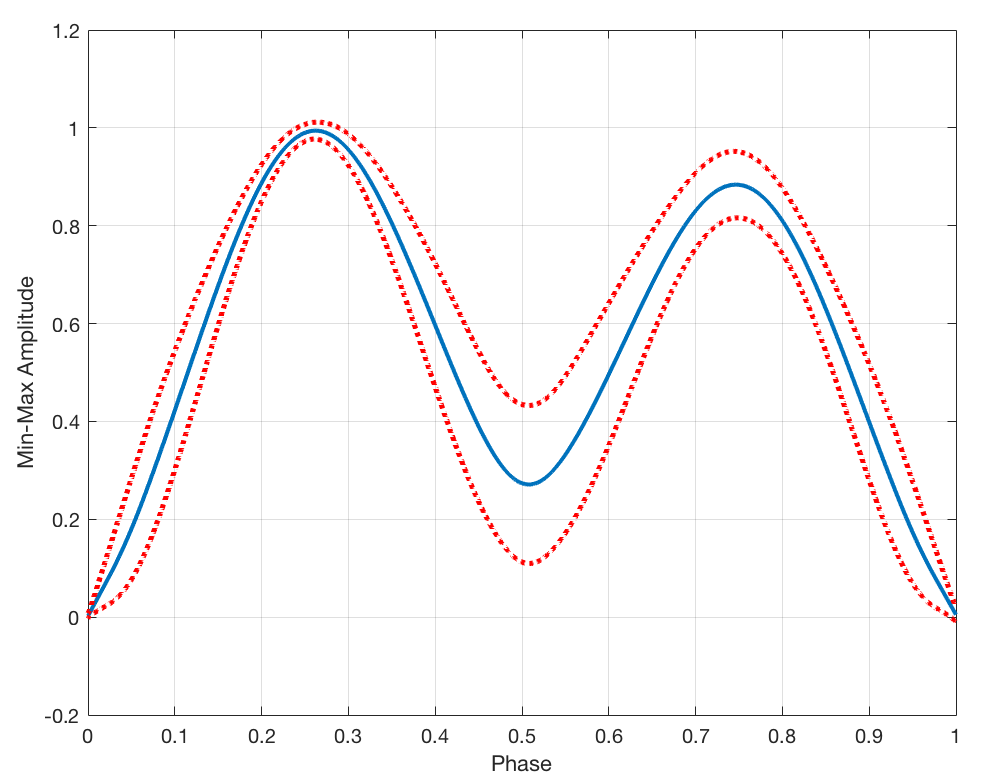}
\par\end{centering}
\caption{The mean (solid) and a $1-\sigma$ standard deviation (dashed) of
the distribution of O'Connell effect Eclipsing Binary phased light
curves discovered via the proposed detector out of the Kepler Eclipsing
Binary catalog. \label{fig:Statistical-Analysis-of}}
\end{figure}

Once we remove the discovered targets that were also in the initial training
data, the result is a conservative selection of 124 potential
targets of interest listed in a supplementary digital file at the project
repository.\footnote{https://github.com/kjohnston82/OCDetector/supplement/AnalysisOfClusters.xlsx} We here present an initial exploratory
data analysis performed on the phased light curve data. At a high level,
the mean and standard deviation of the discovered curves are presented in Figure \ref{fig:Statistical-Analysis-of}.

A more in-depth analysis as to the meaning of the distribution
functional shapes is left for future study. Such an effort would include additional 
observations (spectroscopic and photometric additions would be helpful) as well as 
analysis using binary simulator code such as Wilson--Devinney \citep{Prvsa2005}. It is noted that in general, there are some morphological consistencies across the discovered targets:
\begin{enumerate}
\item In the majority of the discovered OEEB systems, the first maximum following
the primary eclipse is greater than the second maximum following the
secondary eclipse.
\item The light curve relative functional shape from the primary eclipse
(minima) to primary maxima is fairly consistent across all discovered
systems. 
\item The difference in relative amplitude between the two maxima does not
appear to be consistent, nor is the difference in relative amplitude
between the minima.
\end{enumerate}

\begin{figure*}
\begin{centering}
\includegraphics[bb=0in 0bp 2278bp 904bp,scale=0.25]{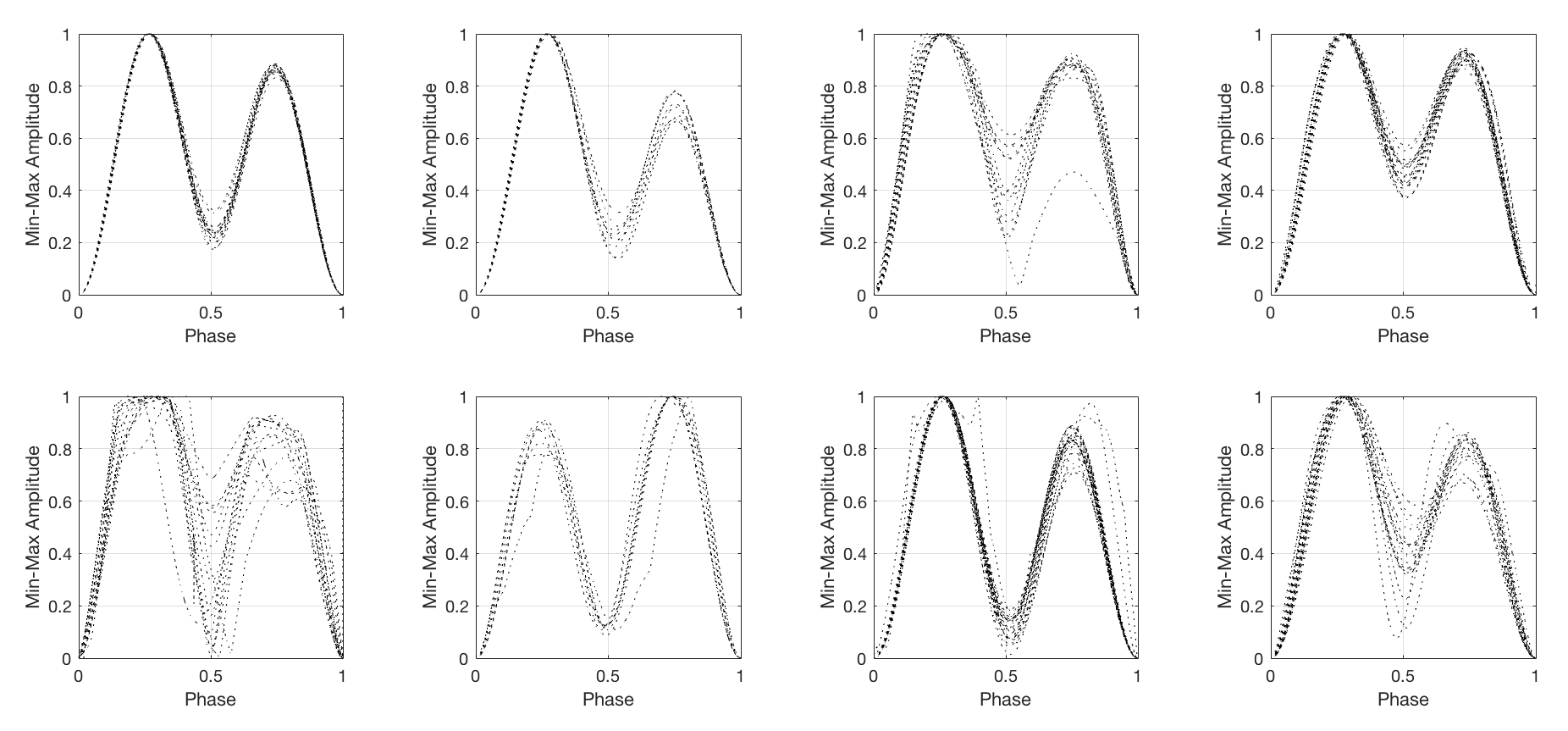}
\par\end{centering}
\caption{The phased light curves of the discovered OEEB data from Kepler, clustered
via k-mean applied to the DF feature space. Cluster number used is based on trial and error,
and the unsupervised classification has been implemented here only to highlight morphological 
similarities. The top four plots represent clusters 1--4 (left to right), and the bottom 
four plots represent clusters 5--8 (left to right). \label{fig:The-clustering-of}}
\end{figure*}

We perform additional exploratory data analysis on the discovered
group via subgrouping partitioning with unsupervised clustering.
The k-means clustering algorithm with matrix-variate distances presented
as part of the comparative methodologies is applied to the discovered
data set (their DF feature space). This clustering is presented to
provide more detail on the discovered group morphological shapes.
The associated 1-D curve generated by the SUPERSMOOTHER algorithm
is presented with respect to their respective clusters (clusters 1--8) in Figure \ref{fig:The-clustering-of}.

\begin{figure}
\begin{centering}
\includegraphics[scale=0.35]{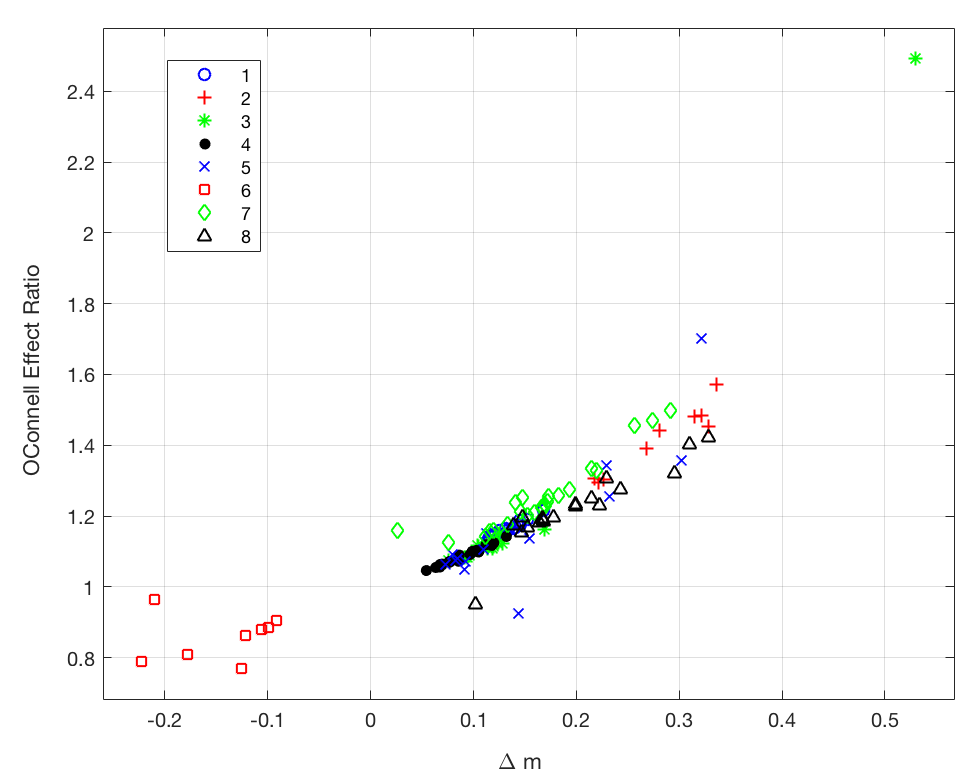}
\par\end{centering}
\centering{}\caption{OER versus $\Delta$m for discovered Kepler O'Connell effect Eclipsing
Binaries. This relationship between OER and $\Delta$m was also demonstrated in \citep{McCartney1999}. \label{fig:OER-vs.-Maximum}}
\end{figure}

The clusters generated were initialized with random starts, thus additional
iterations can potentially result in different groupings. The calculated metric values and the
clusters numbers for each star are presented in the supplementary digital
file. A plot of the measured metrics as well as estimated values of period
and temperature (as reported by the Villanova Kepler Eclipsing Binary
database), are given with respect to the cluster assigned by k-means.\footnote{https://github.com/kjohnston82/OCDetector/Documentation/Figures/ReducedFeaturesKeplerAll\_Temp.png} Following
figure 4.6 in \citep{McCartney1999}, plot of OER versus $\Delta$m
is isolated and presented in Figure \ref{fig:OER-vs.-Maximum}.

We note, that based on our methodology, the selection of initial training data (i.e. the expert selected dataset) will have a strong effect on the performance of the detector and the targets discovered. Biases in the initial training data---limits of $\Delta$m for example---will be reflected in the the discovered dataset. Between our selection of classifier, which is rooted in leveraging similarity with respect to the initial selected training set, and the anomaly detection algorithm we have supplemented our design with, this design effectively ensures that the targets discovered will be limited to morphological features similar to the initial training data provided by the expert. If there were an OEEB with a radically different shape than those stars used in the initial training data, this methodology would likely not find those stars. An increase in missed detection rate has thus been traded for a decrease in false alarm rate, a move that we feel is necessary given the goals of this design and the amount of potential data that could be fed to this algorithm.

\subsubsection{Subgroup Analysis}

The linear relationship between OER and $\Delta m$ reported in
\citep{McCartney1999} is apparent in the discovered Kepler data as
well. The data set here extends from ${\rm OER}\sim(0.7,1.8)$ and $\Delta m\sim\left(-0.3,0.4\right)$,
not including the one sample from cluster 3 that is extreme. This
is comparable to the reported range in \citep{McCartney1999} of ${\rm OER}\sim(0.8,1.2)$
and $\Delta m\sim\left(-0.1,0.05\right)$---a similar OER range, but our
Kepler data span a much larger $\Delta m$ domain, likely resulting
from our additional application of min-max amplitude scaling (i.e., normalized flux). The
gap in $\Delta m$ between $-$0.08 and $0.02$ is caused by the bias
in our training sample and algorithm goal: we only include O'Connell
effect binaries with a user-discernible $\Delta m$. 

The clusters identified by the k-mean algorithm applied to the DF feature space
roughly correspond to groupings in the ${\rm OER}/\Delta$m feature space
(clustering along the diagonal). The individual cluster statistics
(mean and relative error) with respect to the metrics measured here
are given in Table \ref{tab:Example-O'Connell-Effect}. All of the
clusters have a positive mean $\Delta m$, save for cluster 6. The
morphological consistency within a cluster is visually apparent in
Figure \ref{fig:The-clustering-of} but also in the relative
error of LCA, with clusters 5 and 7 being the least consistent. The
next step will include applications to other surveys. 

\begin{table*}
\caption{Metric Measurements from the Discovered O'Connell Effect Eclipsing
Binaries from the Kepler Data Set\label{tab:Example-O'Connell-Effect}}
\begin{tabular*}{\textwidth}{@{\extracolsep\fill}lccccccc}
\hline 
Cluster & $\Delta m$ & $\nicefrac{\sigma_{\Delta m}}{\Delta m}$ & OER & $\nicefrac{\sigma_{\rm OER}}{\rm OER}$ & LCA & $\nicefrac{\sigma_{\rm LCA}}{\rm LCA}$ & $\#$\\ 
\hline 
1 & 0.13 & 0.11 & 1.16 & 0.02 & 7.62 & 0.25 & 17\tabularnewline
2 & 0.28 & 0.17 & 1.41 & 0.07 & 8.92 & 0.16 & 9\tabularnewline
3 & 0.14 & 0.78 & 1.20 & 0.30 & 7.13 & 0.25 & 15\tabularnewline
4 & 0.09 & 0.24 & 1.08 & 0.02 & 6.95 & 0.23 & 22\tabularnewline
5 & 0.15 & 0.55 & 1.17 & 0.16 & 8.54 & 0.58 & 15\tabularnewline
6 & $-$0.14 & $-$0.36 & 0.86 & 0.08 & 8.36 & 0.19 & 8\tabularnewline
7 & 0.17 & 0.36 & 1.25 & 0.08 & 9.41 & 0.82 & 24\tabularnewline
8 & 0.20 & 0.31 & 1.22 & 0.08 & 8.03 & 0.36 & 19\tabularnewline
\hline 
\end{tabular*}

{\it Note.} Metrics are based on \citep{McCartney1999} proposed 
values of interest.
\end{table*}

\subsection{LINEAR Catalog\label{subsec:Cross-Survey-Application}}

We further demonstrate the algorithm with an application
to a separate independent survey. Machine learning methods have been
applied to classifying variable stars observed by the LINEAR
survey \citep{Sesar2011}, and while these methods have focused on
leveraging Fourier domain coefficients and photometric measurements
$\left\{ u,g,r,i,z\right\}$ from SDSS, the data also include best
estimates of period, as all of the variable stars trained on had cyclostationary
signatures. It is then trivial to extract the phased light curve for
each star and apply our Kepler trained detector to the data to generate 
``discovered'' targets of interest.

\begin{table}[H]
\caption{Discovered OEEBs from LINEAR\label{tab:LINEAR_OEEB}}
\centering{}%
\begin{tabular}{cccccccc}
\hline 
13824707 & 19752221 & 257977  & 458198  & 7087932 & 4306725 & 23202141 & 15522736' \tabularnewline
1490274  & 21895776 & 2941388 & 4919865 & 8085095 & 4320508 & 23205293 & 17074729' \tabularnewline
1541626  & 22588921 & 346722  & 4958189 & 8629192 & 6946624'	\tabularnewline
\hline 
\end{tabular}
\end{table}

\begin{figure}
\begin{centering}
\includegraphics[scale=0.35]{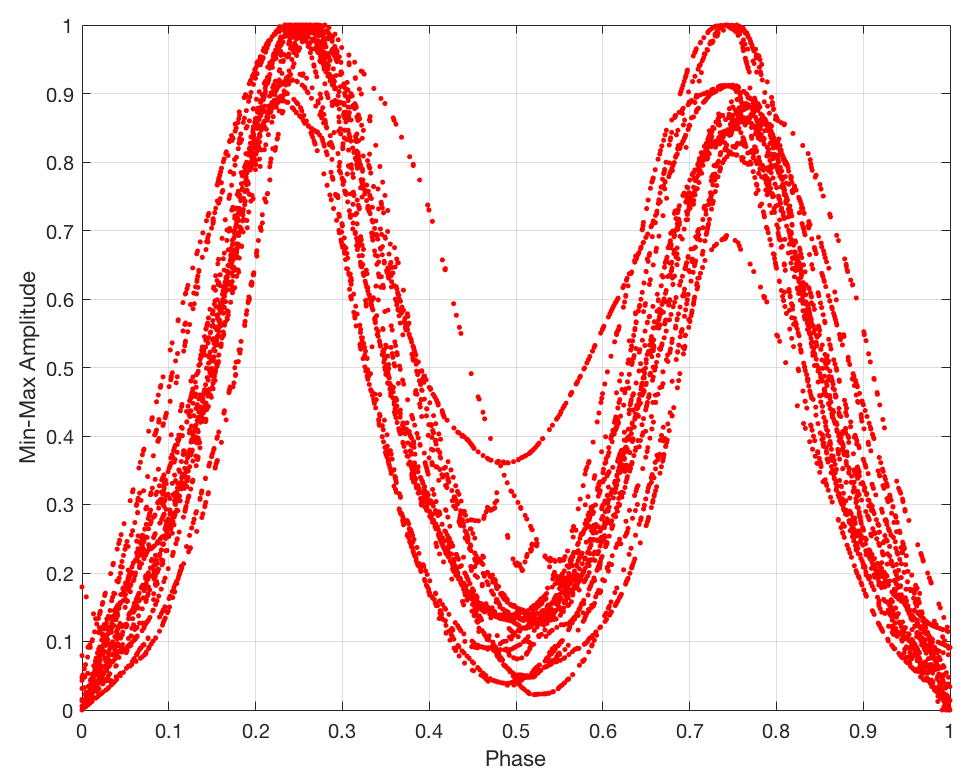}
\par\end{centering}
\caption{Distribution of phased--smoothed light curves from the set of discovered LINEAR 
targets that demonstrate the OEEB signature. LINEAR targets were discovered using the Kepler trained detector. \label{fig:The-discovered-OEEB}}
\end{figure}

The discovered targets are aligned, and the smoothed light curves are presented in
Figure \ref{fig:The-discovered-OEEB}. Note that the LINEAR IDs are presented in Table \ref{tab:LINEAR_OEEB} and as a supplementary digital file at the project repository.\footnote{https://github.com/kjohnston82/OCDetector/supplement/LINEARDiscovered.xlsx}

\begin{figure}
\begin{centering}
\includegraphics[scale=0.35]{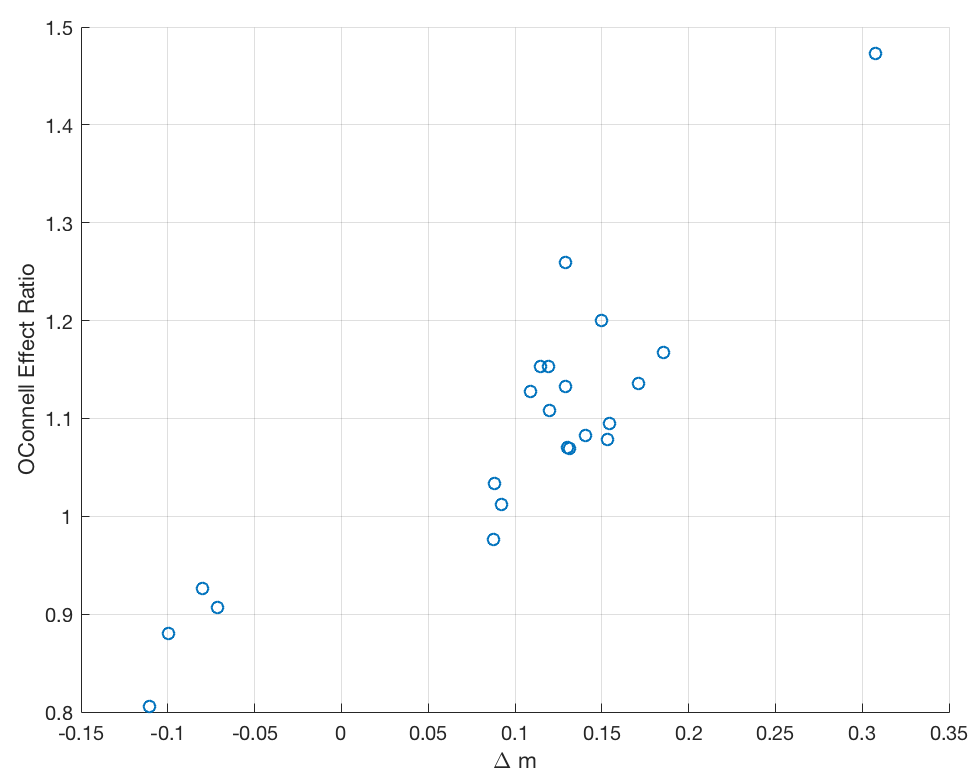}
\par\end{centering}
\caption{OER versus $\Delta$m for the discovered OEEB in the LINEAR data set.
This relationship between OER and $\Delta$m was also demonstrated in \citep{McCartney1999} 
and is similar to the distribution found in Figure \ref{fig:OER-vs.-Maximum}.
\label{fig:OER-vs.-m}}
\end{figure}

Application of our Kepler trained detector to LINEAR data results in 24 ``discovered'' OEEBs. 
These include four targets with a negative O'Connell effect. Similar to the Kepler discovered 
data set, we plot ${\rm OER}/\Delta m$ features using lower-resolution phased binnings ($n=20$) and see
that the distribution and relationship from \citep{McCartney1999}
hold here as well (see Figure \ref{fig:OER-vs.-m}).

\section{Discussion on the Methodology}

\subsection{Comparative Studies\label{subsec:Comparative-Methodologies}}

The pairing of DF feature space and push--pull matrix metric learning
represents a novel design; thus it is difficult to draw conclusions
about performance of the design, as there are no similar studies that
have trained on this particular data set, targeted this particular
variable type, used this feature space, or used this classifier. As we discussed earlier, 
classifiers that implement matrix-variate features directly are few and far between
and almost always not off the shelf. We have developed here two hybrid designs---off-the-shelf 
classifiers mixed with feature space transform---to provide context and comparison. 

These two additional classification methodologies implement
more traditional and well-understood features and classifiers: k-NN using $L^2$ distance 
applied to the phased light curves (Method A) and k-means representation
with quadratic discriminant analysis (QDA) (Method B). Method A is
similar to the UCR \citep{chen2015ucr} time series data baseline
algorithm, reported as part of the database. Provided here is a direct
k-NN classification algorithm applied directly to the smoothed,
aligned, regularly sampled phased light curve. This regular sampling
is generated via interpolation of the smoothed data set and is required
because of the nature of the nearest neighbor algorithm requiring
one-to-one distance. Standard procedures can then be followed \citep{Hastie2009}. 
Method B borrows from \cite{park2003lower}, transforming the matrix-variate data 
into vector-variate data via estimation of distances between our training 
set and a smaller set of exemplar means DFs that were generated via unsupervised
learning. Distances were found using the Frobenius norm of the difference between the two
matrices.

Whereas Method A uses neither the DF feature representation nor the
metric learning methodology, Method B uses DF feature space but not
the metric learning methodology. This presents a problem, however,
as most standard out-of-the-box classification methods require a vector
input. Indeed, many methodologies, even when faced with a matrix input,
choose to vectorize the matrix. An alternative to this implementation
is a secondary transformation into a lower-dimensional feature space.
Following the work of \citep{park2003lower}, we implement a matrix distance k-means
algorithm (e.g., k-means with a Frobenius norm) to
generate estimates of clusters in the DF space. Observations are transformed
by finding the Euclidean distance between each training point and each
of the k-mean matrices discovered. The resulting set of k-distances
is treated as the input pattern, allowing the use of the standard
QDA algorithm \citep{Duda2012}. The performances of both the
proposed methodology and the two comparative methodologies are presented
in Table \ref{tab:PerformanceEstimates-1}. The algorithms are available
as open source code, along with our novel implementation, at the project
repository.

\begin{table}
\caption{Comparison of Performance Estimates across the Proposed Classifiers
(Based on Testing Data) \label{tab:PerformanceEstimates-1}}
\centering{}%
\begin{tabular*}{\textwidth}{@{\extracolsep\fill}lccc}
\hline 
 & PPML & Method A & Method B\tabularnewline
\hline 
Error rate & 12.5\% & 15.6\% & 12.7\%\tabularnewline
\hline 
\end{tabular*}
\end{table}

We present the performance of the main novel feature space/classification
pairing as well as the two additional implementations that rely on
more standard methods. Here we have evaluated performance based on 
misclassification rate, i.e., 1-accuracy given by \cite{fawcett2006} as $1 - {\rm correct}/{\rm total}$. 
The method we propose has a marginally better
misclassification rate (Table \ref{tab:PerformanceEstimates-1})
and has the added benefit of (1) not requiring unsupervised clustering,
which can be inconsistent, and (2) providing nearest neighbor estimates
allowing for demonstration of direct comparison. These performance
estimate values are dependent on the initial selected training and
testing data. They have been averaged and optimized via cross-validation;
however, with so little initial training data and with the selection
process for which training and testing data are randomized, performance
estimates may vary. Of course, increases in training data will result
in increased confidence in performance results.

We have not included computational times as part of this analysis, as they tend to be dependent
on the system operated on. We can anecdotally discuss that, on the system implemented as part
of this research (MacBook Pro, 2.5 GHz Intel i7, 8 GB RAM), the training optimization of our 
proposed feature extraction and PPML classification total took less than 5--10 min to run---variation 
depending on whatever else was running in the background. Use of the classifiers on unlabeled data 
resulted in a classification in fractions of seconds per star. However, we should note that this 
algorithm will speed up if it is implemented on a parallel processing system, as much of the time
taken in the training process resulted from linear algebra operations that can be parallelized.

\subsection{Strength of the Tools}

The DF representation maps deterministic,
functional stellar variable observations to a stochastic matrix, with
the rows summing to unity. The inherently probabilistic nature of
DFs provides a robust way to model interclass variability and handle
irregular sampling rates associated with stellar observations. Because the DF feature is 
indifferent to sampling density so long as all points along the functional 
shape are represented, the trained detection algorithm we generate and demonstrate in this article can
be trained on Kepler data but directly applied to the LINEAR data, as shown in
section \ref{subsec:Cross-Survey-Application}. 

The algorithm, including comparison methodologies, designed feature
space transformations, classifiers, utilities, and so on, is publicly available
at the project repository;\footnote{https://GitHub.com/kjohnston82/OCDetector}
all code was developed in MATLAB and was run on MATLAB 9.3.0.713579
(R2017b). The operations included here can be executed either via calling individual functions
or using the script provided (ImplementDetector.m). Likewise, a Java version of all of the individual computational functions has been generated \citep[see JVarStar, ][]{Johnston2019ascl.soft04029J}
and is included in the project repository.\footnote{https://GitHub.com/kjohnston82/VariableStarAnalysis}

\subsection{Perspectives}

This design is modular enough to be applied as is to other types of stars and star systems that are cyclostationary in nature. With a change in feature space, specifically one that is tailored to the target signatures of interest and based on prior experience, this design can be replicated for other targets that do not demonstrate a cyclostationary signal (i.e., impulsive, nonstationary, etc.) and even to targets of interest that are not time variable in nature
but have a consistent observable signature (e.g., spectrum, photometry, image point-spread function, etc.). One of the advantages of attempting to identify the O'Connell effect Eclipsing Binary is that one only needs the phased light curve---and thus the dominant period allowing a  phasing of the light curve---to perform the feature extraction and thus the classification. The DF process here allows for a direct transformation into a singular feature space that focuses on functional shape.

For other variable stars, a multiview approach might be necessary; either descriptions of the light curve signal across multiple transformations (e.g., Wavelet and DF), or across representations (e.g. polarimetry and photometry) or across frequency regimes (e.g. optical and radio) would be required in the process of properly defining the variable star type. The solution to this multiview problem is neither straightforward nor well understood \citep{akaho2006kernel}. Multiple options have been explored to resolve this problem: combination of classifiers, canonical correlation analysis, postprobability blending, and multimetric
classification. The computational needs of the algorithm have only been roughly studied, and a more thorough review is necessary in the context of the algorithm proposed and the needs of the astronomy community. The k-NN algorithm dependence on pairwise difference, while one
of its strong suits is also one of the more computationally demanding parts of the algorithm. Functionality such as $k-d$ trees
as well as other feature space partitioning methods have been shown to reduce the computational requirements. 

\section{Conclusion}

The method we have outlined here has demonstrated the ability to detect targets of interest given a training set consisting of expertly labeled light curve training data. The procedure presents two new functionalities: the distribution field, a shape-based feature space, and the push--pull matrix metric learning algorithm, a metric learning algorithm derived
from LMNN that allows for matrix-variate similarity comparisons. As a demonstration, the design is applied to Kepler eclipsing
binary data and LINEAR data. The methodology proposed---DF +
Push-Pull Metric Learning---is comparable to other methods presented, with respect to the OEEB detection problem given the limited Kepler dataset we have used for training. Furthermore, the increase in the number
of systems, and the presentation of the data, allows us to make additional
observations about the distribution of curves and trends within the
population. Future work will involve the analysis of these statistical
distributions, as well as an inference as to their physical meaning.

The new OEEB systems we discovered by the method of automated detection proposed here can be used to further investigate their frequency of occurrence, provide constraints on existing light curve models, and provide parameters to look for these systems in future large-scale variability surveys like LSST. Although the effort here targets OEEB as a demonstration, it need not be limited to those
particular targets. We could use the DF feature space along with the push--pull metric learning classifier to construct a detector for any variable stars with periodic variability.
Furthermore, any variable star (e.g., supernova, RR Lyr, Cepheids, eclipsing binaries) can be targeted using this classification scheme, given the appropriate feature
space transformation allowing for quantitative evaluation of similarity. This design is directly applicable to exo-planet discovery; either via light curve detection (e.g., to detect eclipsing exo-planets) or via machine learning applied to other means (e.g., spectral analysis). 

\bibliography{OCDetector}

\bibliographystyle{Science}


\section*{Abbreviations}
TESS, Transiting Exoplanet Survey; KELT, Kilodegree Extremely Little Telescope; LSST, Large Synoptic Survey Telescope; DF, distribution field; OEEBs, O'Connell effect eclipsing binaries; LINEAR, Lincoln Near-Earth Asteroid Research ; MATLAB, Matrix Laboratory; QDA, quadratic discriminate analysis; UCR, University of California Riverside; OER, O'Connell effect ratio; LCA, light curve asymmetry; SDSS, Slone Digital Sky Survey 

\section*{Availability of data and materials}
The software developed for this article, as well as the reduced data results, is available at https://GitHub.com/kjohnston82/OCDetector. The links to the raw training data are also provided at the public repository.

\section*{Competing interests}
 The authors declare that they have no competing interests.
 
\section*{Funding}
Research was partially supported by Perspecta, Inc. This material is based upon work supported by the NASA Florida Space Grant under 2018 Dissertation and Thesis Improvement
Fellowship No. 202379. The LINEAR program is sponsored by the National Aeronautics and Space Administration (NRA Nos. NNH09ZDA001N, 09-NEOO09-0010) and the U.S. Air Force under Air Force contract FA8721-05-C-0002.

\section*{Author's contributions}
The detector code was developed by KBJ and RH. Initial targets of interest were identified by MK. Training data and testing data were procured and refined by KBJ. KBJ prepared this manuscript and performed the tests detailed within. All authors read and approved the final manuscript.

\section*{Acknowledgements}
The authors are grateful to the anonymous reviewers for their detailed and expansive recommendations on the organization of this article. The authors are grateful for the valuable astrophysics
insight provided by C. Fletcher and T. Doyle. Inspiration provided by C. Role. Initial editing and
review provided by H. Monteith. The authors would like to additionally acknowledge the Kepler Eclipsing Binary team, without whom much of the training data would not exist.

\end{document}